\documentstyle[tighten,preprint,amsfonts,%
epsf,eqsecnum,aps,prd]{revtex}
\begin{document}
\draft

\font\tenbf=cmbx10 
\font\tenrm=cmr10 
\font\tenit=cmti10 
\font\elevenbf=cmbx10 scaled\magstep 1 
\font\elevenrm=cmr10 scaled\magstep 1 
\font\elevenit=cmti10 scaled\magstep 1 
\font\ninebf=cmbx9 
\font\ninerm=cmr9 
\font\nineit=cmti9
\font\eightbf=cmbx8 
\font\eightrm=cmr8 
\font\eightit=cmti8 
\font\sevenbf=cmbx7
\font\sevenrm=cmr7 
\font\sevenit=cmti7 
\font\elevensl=cmsl10 scaled\magstep 1 \font\eleveni=cmmi10 scaled\magstep 1
\font\elevensy=cmsy10 scaled\magstep 1 \font\elevenex=cmex10 scaled\magstep1
\font\elevensc=cmcsc10 scaled\magstep1 
\skewchar\eleveni='177 \skewchar\elevensy='60

\font\seveni=cmmi7 
\font\sevensy=cmsy7

\def\scri{\hbox{${\cal J}$\kern -.645em {\raise .57ex\hbox{$\scriptscriptstyle (\
$}}}} 
\def\frac#1#2{{#1 \over #2}} 
\narrowtext
\vglue 1cm 
\hsize=6.0truein 
\vsize=8.5truein 
\hoffset=0.325truein
\voffset=0.375truein
\parindent=3pc 
\baselineskip=10pt
\centerline{\elevenbf TOPOLOGICAL CENSORSHIP} 
\vglue 1.0cm 
\centerline{\tenrm KRISTIN SCHLEICH } 
\vglue 0.3cm 
\centerline{\tenrm and} 
\vglue 0.3cm
\centerline{\tenrm DONALD M. WITT} 
\baselineskip=13pt 
\centerline{\tenit Department of Physics, University of British Columbia} 
\baselineskip=12pt
\centerline{\tenit Vancouver, BC V6T 1Z1, Canada} 
\vglue 0.8cm 
\centerline{\tenrm ABSTRACT} 
\vglue 0.3cm 
{\tenrm\baselineskip=12pt 
\noindent Classically, all
topologies are allowed as solutions to the Einstein equations.  However,  one
does not observe any topological structures on medium range distance scales, that
is scales that are smaller than the size of the observed universe but larger than
the microscopic scales for which quantum  gravity becomes important.  Recently,
Friedman, Schleich and Witt have proven that there is topological censorship on
these medium range distance scales: the Einstein equations, locally positive
energy, and local predictability of physics imply that any medium distance scale
topological structures  cannot be seen.  More precisely, we show that the
topology of physically reasonable isolated systems is shrouded from distant
observers, or in other words there is a {\tenit topological censorship
principle}. } 
\narrowtext
\vglue 0.8cm 
\noindent{\elevenbf 1. Introduction} 
\vglue 0.4cm
 
\baselineskip=14pt 

An interesting observed fact about our universe
is that its spatial topology is trivial,  that is continuously deformable to
three dimensional Euclidean space, that is a region of ${\bf R}^3$. Thus is true on a 
remarkably large range of 
distance scales, ranging from fermis, the scale of interactions of high energy
particles to megaparsecs, the scale of intergalactic distances. The theory of general 
relativity does not select a trivial
topology;  all 3-manifolds occur as the spatial topology of solutions to the
Einstein equations.$^1$ Thus general relativity allows spacetimes that contain
arbitrarily large numbers of wormholes and other complicated topological
structures. In fact, there even exist inflationary universes with an arbitrarily
wide range of topology.$^2$ Moreover, the dynamics of general relativity do not
allow for the topology to change. Why don't we see topological structures in our
spacetime?

At first the answer to this question might seem obvious; we see no topological
structures because they are not there. However, there is a more interesting
possibility; such topological structures may indeed be present in our universe,
but cannot be observed. This second possibility was more formally stated as the
{\elevenit topological censorship conjecture} by Friedman and Witt: The topology
of any physically reasonable isolated system is shrouded. That is a spacetime may
contain isolated topological structures such as wormholes; however, there is no
method by which an experimenter can determine that a spacetime contains such
non-Euclidean topology and report this result to a distant observer. 

Recently Friedman, Schleich and Witt have proven this conjecture.$^3$ Today,
instead of emphasizing the technical details used in proving the result, I would
like to give an informal and intuitive sketch of how this proof works and mention
some of its consequences for general relativity and cosmology.

\vglue 0.6cm 
\noindent{\elevenbf 2. Background} 
\vglue 0.4cm

One of the fundamental tenets of classical physics is {\elevenit physical
predictability}, that is given knowledge of the initial condition of the system,
the laws of physics allow one to determine the behavior of the system at all
future (and past) times. For example the motion of a particle in a potential is
physically predictable; given the position and velocity of the particle at one
instant of time, the equations of motion when solved determine the trajectory of
the particle for all time. Similarly a spacetime is physically predictable if
information about its geometry and matter sources at one instant of time  allows
the determination of its spacetime geometry for all times via the Einstein equations.
For those familiar with general relativity, physically predictable spacetimes are
termed {\elevenit globally hyperbolic} spacetimes.$^4$  Such a definition is
needed because not all solutions of the Einstein equations are physically
predictable. For example, solutions that contain closed timelike curves or naked
singularities are not physically predictable; however such solutions exhibit properties
that patently violate the laws of physics.
Thus  it is the  properties of physically predictable spacetimes
that are of great interest as our universe is of this type.

Topology describes the properties of a space that are independent of the metric,
that is the distances between points in the space. A familiar two dimensional
example is given by the torus.
Observe that there is one hole in this space;  moreover, this hole remains under
any continuous deformation of the surface, that is under any continuous twisting
or stretching of the surface. A sphere has no hole; therefore, there is no continuous 
deformation of
the torus that takes it into a sphere. This observation is formally stated in
terms of topology: a sphere and a torus have different topology. 
This difference in topology
can be seen in the properties of curves on the torus and the
sphere: One can find closed curves on the  torus that are noncontractible, that
is they cannot be continuously shrunk to a point. Such curves are those that loop
about the hole in the torus. However, all closed curves on the sphere are
contractible. In fact, the properties of curves on the two spaces rigorously
characterize their
topology. 

How would an experimenter go about determining the topology of spacetime? An
important fact to remember is that experimenters, like all other massive objects,
travel on timelike paths through their spacetime. Moreover, any measurement that
an experimenter performs relies on information that also travels on a timelike or
null path. Note especially that the path need not be everywhere timelike or
everywhere null; for example, information could be carried from one point to
another first by a massive particle travelling a timelike path that then
releases a photon travelling a null path. \vadjust{\eject} Thus a determination of the topology of
spacetime must be carried out using information that travels  a path that is
either timelike or null at all points along its course. Such paths are referred
to as {\elevenit causal curves}.\footnote{\hsize=6.0truein
\ninerm\baselineskip=11pt An observant reader
may worry that our universe is not two dimensional, but four dimensional;
however, properties of curves in it still characterize its topology. Our universe
is a physically predictable spacetime and thus its topology is completely carried
by the topology of the three dimensional spatial hypersurface; physical
predictability implies the topology at one time must determine it for all times.
The topology of a three dimensional hypersurface can be characterized finding
noncontractible curves as in the two dimensional case. Therefore, by laying out
causal curves in the four dimensional spacetime, an experimenter can actively
probe its topology.}

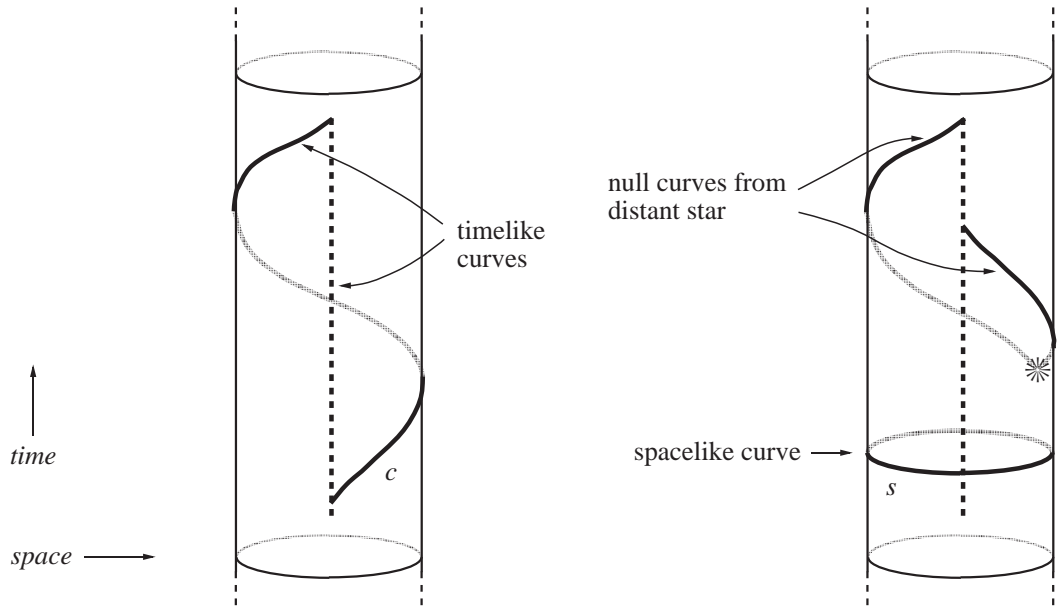
\begin{center}
\vglue .5 cm
\leavevmode
\epsfysize=8cm
\epsfbox{cylspace.epsf}
\end{center}

\vskip 0.20in
{\baselineskip = 12pt \tenrm \noindent
1. A two dimensional spacetime with
nontrivial topology. An observer can detect this topology through
nontrivial timelike curves such as {\tenit c} as shown on the left, or through
 nontrivial null
curves, as shown on the right. The observer cannot detect the topology using a
spatial curve, such as {\tenit s} as information cannot travel such a curve.} 
\vskip 0.20in

 For example, an experimenter living in the two dimensional spacetime
illustrated in figure 1 can  determine that the spatial topology is a
circle by discovering that there are noncontractible causal curves in the spacetime. 
The
experimenter can find such  curves, for example, by laying out a rope while walking
around the space counterclockwise as illustrated by curve $c$ in figure 1. The
resulting loop of rope clearly will not be contractible. Note that the
experimenter cannot use a noncontractible spacelike curve such as $s$
to determine the topology as it is physically impossible to lay out such a curve.
It is especially important to note that the experimenter can probe the topology
using distant objects that emit information such as massive particles or light 
that travels
along causal curves.
For example the experimenter may be able to use light travelling two distinct paths
from  a star to probe the topology as also illustrated in figure 1.

It is apparent that the causal structure of spacetime, that is which points of the
spacetime can be connected to each other by causal curves, is important to the
topological censorship theorem; after all the question at hand  is whether or not
causal curves can thread the topology and come out at any time, even infinitely
far in the future. Therefore, a concrete way of talking about such curves is
needed. This terminology and a very useful pictorial way of representing this
causal structure, Penrose diagrams, is based off of the causal structure of
Minkowski space.\footnote{\hsize=6.0truein\ninerm\baselineskip=11pt  
Ref.~4, p.~118 provides a
clear discussion of  terminology used in causal structure and Penrose diagrams as
well as a comprehensive discussion of many of the other concepts used in the
theorem. References to the original literature also can be found in this text.\hfil}

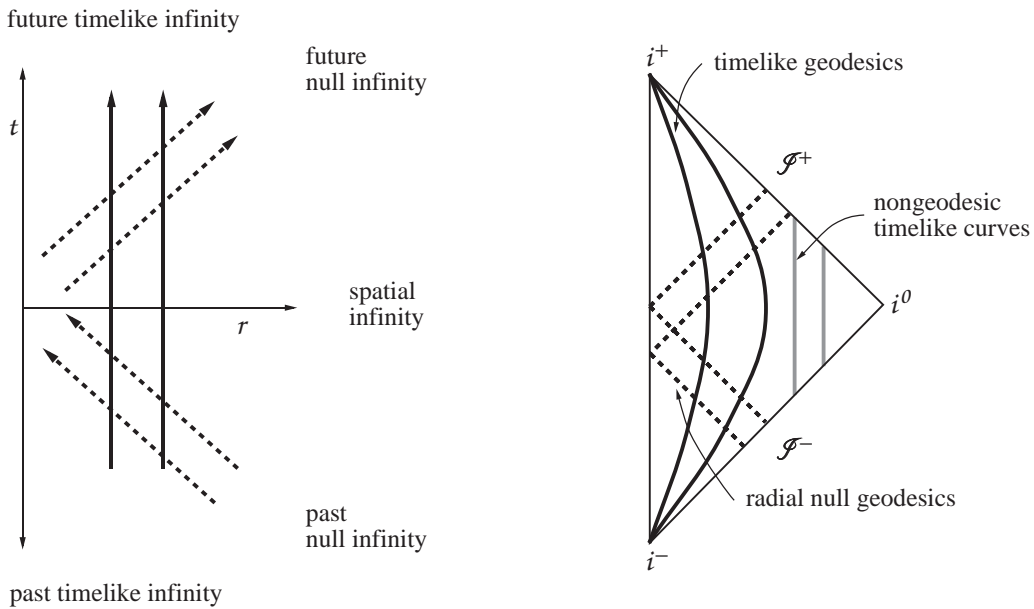
\begin{center}
\vglue .5 cm
\leavevmode
\epsfysize=8cm
\epsfbox{penrose.epsf}
\end{center}

\vskip 0.20in
{\baselineskip = 12pt \tenrm \noindent 2. The causal structure of Minkowski
spacetime. On the left, Minkowski spacetime is represented in spherical
coordinates as a two dimensional diagram by suppressing the angular coordinates.
On the right is the Penrose diagram for Minkowski space. Points once off
at infinite distance such as future and past timelike infinity and future and
past null infinity now appear at finite distance.}

\vskip 0.20in

An illustration of Minkowski spacetime with metric 
$$ds^2 = -dt^2 + dr^2
+r^2(d\theta^2 + \hbox{\elevenrm sin}^2\theta d\phi^2)\eqno(1)$$ 
is given in the left side  of
figure 2. Note that there is no concrete way in this diagram to represent where
the timelike geodesics or the null geodesics end up at in the infinite future.
For example, it is not clear from this diagram that null geodesics go to to a
different  infinity than timelike geodesics. The reason is obvious; the
coordinates $r$, and $t$ have infinite ranges; therefore 
there is no way to draw where infinity is. However, this
is not a problem with  studying causal structure; rather it is a problem with the
choice of coordinates. But physics is coordinate invariant,  so instead write 
the metric of Minkowski
space in a set of coordinates such that these infinite points now occur at finite
coordinate values. Defining $t'$ and $r'$ by 
$$\hbox{\elevenrm 2}t = \hbox{\elevenrm tan} (\frac{t'+r'}2)
+\hbox{\elevenrm tan}(\frac{t'-r'}2) \ \ \ \ \ \ \ \ \ \hbox{\elevenrm 2}r= 
\hbox{\elevenrm tan} (\frac{t'+r'}2)
-\hbox{\elevenrm tan}(\frac{t'-r'}2)$$ 
the Minkowski metric becomes
$$ds^2=\Omega^2(t',r')\biggl(-dt'^2 +dr'^2 + r'^2(d\theta + 
\hbox{\elevenrm sin}^2\theta
d\phi^2)\biggr)\eqno(2)$$ 
where $\Omega(t',r') = \frac 12 \hbox{\elevenrm sec}(\frac{t'+r'}2)
\hbox{\elevenrm sec} (\frac{t'-r'}2)$. The new coordinates have finite ranges, 
$r'\ge
0$, $-\pi < t'+r' <\pi$, and $-\pi < t'-r' <\pi$; thus the full spacetime can now
be represented in a finite diagram. Finally, one uses the fact that two
spacetimes related by a conformal transformation have the same causal structure
to further simplify the discussion; this fact means that one need not represent
the factor of $\Omega^2$ in the above metric to concretely illustrate the causal
structure. The resulting diagram, the Penrose diagram of Minkowski spacetime is
given in the right side of figure 2. By changing coordinates, the infinite
future and past are now clearly described. Timelike geodesics, for example the
path travelled an observer who remains at a fixed radial coordinate position $r$
begin at {\elevenit past timelike infinity}, $i^-$, and end at {\elevenit future
timelike infinity}, $i^+$. Similarly null geodesics, for example the path
travelled by a photon travelling radially inward to the coordinate origin begin
at {\elevenit past null infinity},  $\scri^-$, and end at {\elevenit future null
infinity}, $\scri^+$. It is now clear that the future infinities of that null
geodesics and timelike geodesics are distinct. Observe that not all timelike curves
end at $i^+$; curves corresponding to accelerated timelike observers can reach
$\scri^+$. Finally, radially directed photons travel along paths at 45
degree angles in this spacetime; thus any timelike or null curve leaving a point
in this spacetime must have tangent lying between the inward directed radial null
geodesic and outward directed radial null geodesic. Therefore, this Penrose
diagram neatly encapsulates the information about the causal structure of
Minkowski spacetime. 

It is clear that technique used to concretely discuss the causal
structure of Minkowski spacetime can be applied to
illustrate the causal structure of other spacetimes. Of course, spacetimes with
different metrics and topology will not have Penrose diagrams that are identical
to figure 2. However, spacetimes that resemble or approach Minkowski space in
regions of the spacetime will have similar causal structure in those regions. In
particular, spacetimes containing isolated topological structures will have
similar causal structure far away from the topology.

An isolated topological structure is, as implied by its name, one that can be
isolated from the rest of the spacetime. More precisely, one can place a sphere
around the topology at a given instant in time and "cut out" the topology by
excising this sphere and everything inside it from the spatial hypersurface. When
one does so, the remaining space has the topology of three dimensional Euclidean
space minus a ball. Note that the sphere surrounding the topology might be very
large and contain many topological structures; for example, there might be a
large number of wormholes inside the sphere. Additionally,  the metric of the
spacetime outside the evolution of the excised ball approaches Minkowski
spacetime as one goes infinitely far away; that is one can find a set of
coordinates such that as one goes to infinite spatial distance, the metric is
 Eq.~(1) plus $1/r$ correction
terms.  A spacetime  that satisfies these conditions is termed an
{\elevenit asymptotically flat spacetime}.  The Schwarzchild solution is a canonical
example of an asymptotically flat spacetime, though it is obvious that there
are myriad examples of such spacetimes, including those with isolated topological
structures.

It is important for understanding the theorem to observe that there may be more
than one asymptotically flat region for a spacetime containing isolated topological
structures; the topological structures  potentially can connect several different
copies of ${\bf R}^3$, each of which admits a metric approaching that of Minkowski
space.\footnote{\hsize=6.0truein\ninerm\baselineskip=11pt   A spatial analog of 
this feature
is provided by figure 4, although it is included in this paper
for other purposes. It contains two  spatially asymptotically flat regions
connected by a throat. \hfil} Additionally note that as the spacetime is
approaching Minkowski spacetime in each asymptotically flat region, intuitively
its causal structure is also approaching that of Minkowski
spacetime. Indeed this is the case; the Penrose diagram for an asymptotically
flat spacetime will contain one or more regions which have the same causal
structure as Minkowski spacetime.
\vglue 0.6cm 
\noindent{\elevenbf 3. The Topological Censorship Theorem} 
\vglue 0.4cm

The Einstein equations can be coupled to a wide range of matter, but certain
general properties characterize classical physical matter, that is sources that
are not quantum in nature. These properties are called energy conditions.
The energy condition used in the proof of the topological censorship theorem is
the {\elevenit null energy condition}.$^6$ Physically, this condition states that an
observer travelling along either a timelike or null curve measures the energy in
their local frame to be positive at any point in the spacetime. This condition is
satisfied by all  classical sources of matter found in nature such as gas, dust,
radiation, electromagnetic fields, as well as idealized sources such as classical
scalar fields. It is implied by each of the other classical energy conditions: the
weak energy condition, the strong energy condition and the dominant energy
condition. Therefore, the null energy condition is a very reasonable and physical
restriction.

Given the above, it is now possible to state the theorem:
\vskip 14pt
\noindent {\elevenbf Theorem.} {\elevensl If an asymptotically flat globally
hyperbolic spacetime satisfies the null energy condition, then every
causal curve from past null infinity to future null infinity is deformable
to $\gamma_0$.}
\vskip 14pt
The curve $\gamma_0$ is a representative causal curve with past endpoint at
$\scri^-$ and future endpoint at $\scri^+$ that lies  in the asymptotically flat
region; i.e. it is a curve that does not pass through any of the topology of the
spacetime.

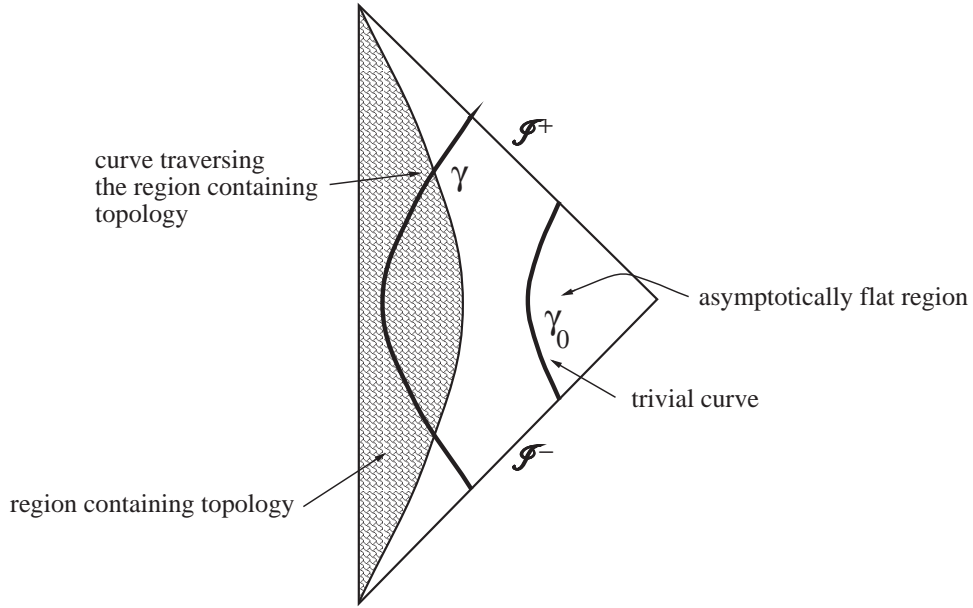
\begin{center}
\vglue .5 cm
\leavevmode
\epsfysize=8cm
\epsfbox{topology1.epsf}
\end{center}

\vskip 0.20in
{\baselineskip = 12pt \tenrm \noindent 3. A Penrose diagram illustrating the
statement of the theorem. The theorem proves that any causal curve  traversing the
topology in the shaded region cannot reach future null infinity. Thus the curve $\gamma$ in
this diagram either does not go through the topology or is spacelike at points along its course.}

\vskip 0.20in

Figure 3 is a Penrose diagram of a spacetime used  to illustrate the
theorem; for convenience this spacetime is assumed to have only one asymptotic
region. Note especially that the causal properties of the shaded region, that is
the region containing topology are not faithfully illustrated; only those of the
asymptotically flat region, that  with
the causal structure of the radially
distant part of Minkowski spacetime, have been correctly diagrammed.
The theorem  states that any causal curve $\gamma$ that can reach an observer in
the asymptotically flat region is deformable to the trivial curve $\gamma_0$.
This means that $\gamma$ cannot traverse any topological structure because if it
did, it could not be deformed into the trivial one as it would hook on the
topology. It follows that all causal curves that enter a topological structure
cannot come out again; therefore there is no way to actively probe  the topology.

The proof of the theorem is based
on a lemma that applies to simply connected spacetimes. A simply connected
spacetime is one for which all closed curves are contractible. For example, a
sphere and a plane are simply connected spacetimes, but a torus is not. Thus
simply connected spacetimes have very special topology.
\vskip 14pt
\noindent {\elevenbf Lemma.} {\elevensl Suppose one has an asymptotically flat
simply connected spacetime that satisfies the null energy condition. Then no
2-surface $\tau$ that is outer trapped with respect to $\scri$ can be seen from
$\scri^+$.}
\vskip 14pt
A surface is said to be {\elevenit outer trapped} if radially outward directed 
null rays are
converging.$^7$ Note that radially outward is defined by the direction one goes to
reach the asymptotically flat region of the observer. In order to understand the
motivation behind this definition, consider the behavior of light emitting
surfaces in both Minkowski and curved spacetime. First take a sphere at one
instant of time in Minkowski space and release radially outward directed photons
from its surface. At some small interval of time later, say one second, consider
the surface formed by these photons. As the spacetime geometry is flat, this
photon sphere will have larger area than that of the original sphere. This result
seems obvious; however, if one works in a curved spacetime the result can be very
different and depends on the curvature of the spacetime in the region of the
sphere. An example is illustrated in figure 4; the surface $\tau$ in this
particular spacetime has a photon sphere that has larger area at later time, 
very similar to
the situation in Minkowski space. However, the surface ${\tau}'$ has a photon
sphere with smaller area at later time; the curvature of the spacetime in the 
neighborhood of
${\tau}'$ forces the light rays to converge even though the begin by heading in
the  radially outward direction. Thus this surface is outer trapped. Physically
what is happening is the curvature of spacetime is so strong that even radially
outward directed light rays are being forced inward.

\begin{center}
\vglue .5 cm
\leavevmode
\epsfysize=6cm
\epsfbox{outertrap.epsf}
\end{center}

\vskip 0.20in
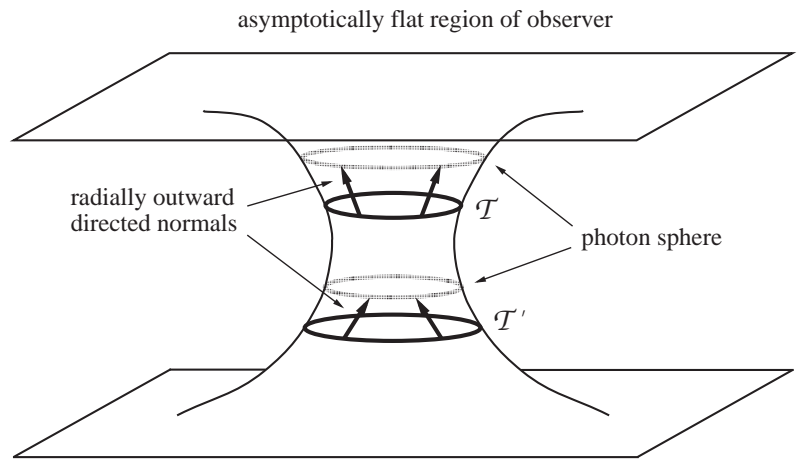
{\baselineskip = 12pt \tenrm \noindent 4. Illustration of an outer trapped
surface. The picture represents a three dimensional spatial slice of a spacetime; 
one dimension is
suppressed.
 $\tau$  and ${\tau}'$ are spatial spheres in the hypersurface.
$\tau$ is not outer trapped; ${\tau}'$ is outer trapped.} 
\vskip 0.20in

Given this physical picture of an outer trapped surface one intuitively gathers
that no information from it or interior to it can escape out to asymptotic
infinity to be detected by the observer; no signal travels faster than light and
radially outward directed light is travelling outward at the maximal rate.
Therefore if radially directed light is forced spatially inward by the curvature,
all other signals, radial or nonradial will be as well. Thus you cannot
see inside a trapped surface. This is precisely what
the Lemma proves rigorously.  I will not go into the proof here, but note that it
uses standard techniques from the singularity theorems.$^5$

This lemma is for a simply connected spacetime, but the theorem applies to all
asymptotically flat spacetimes. The way that a connection is made is through the
concept of a {\elevenit universal cover}. A universal cover is a spacetime that
is related to the original spacetime through properties of curves; curves that
are noncontractible in the original space are unwrapped to become contractible
curves in the universal cover. Again it is easiest to illustrate this concept
through an example. An illustration of the cylinder spacetime and its universal
cover are given in Figure 5. In the original spacetime,  $b$ is a
trivial curve,  $c$, $s$ and $d$ are all nontrivial curves. Its universal
cover is a plane; one can wrap the plane around the cylinder to cover it an
infinite number of times. Each point in the cylinder is covered by  an infinite
number of points in the plane, for example $q$ corresponds to an infinite number
of points in the plane as indicated. Now noncontractible curves in the cylinder
spacetime are identified with contractible curves connecting different copies of
the starting point $q$ to $p$ in the universal covering spacetime. For example
the curve $c$ is identified with a curve attaching one copy of
$q$  to $p$, the curve $d$ is identified with a curve attaching a different copy
of $q$ to $p$. Clearly,  by construction the universal cover is a
simply connected spacetime.

\begin{center}
\vglue .5 cm
\leavevmode
\epsfysize=7cm
\epsfbox{cover.epsf}
\end{center}

\vskip 0.20in
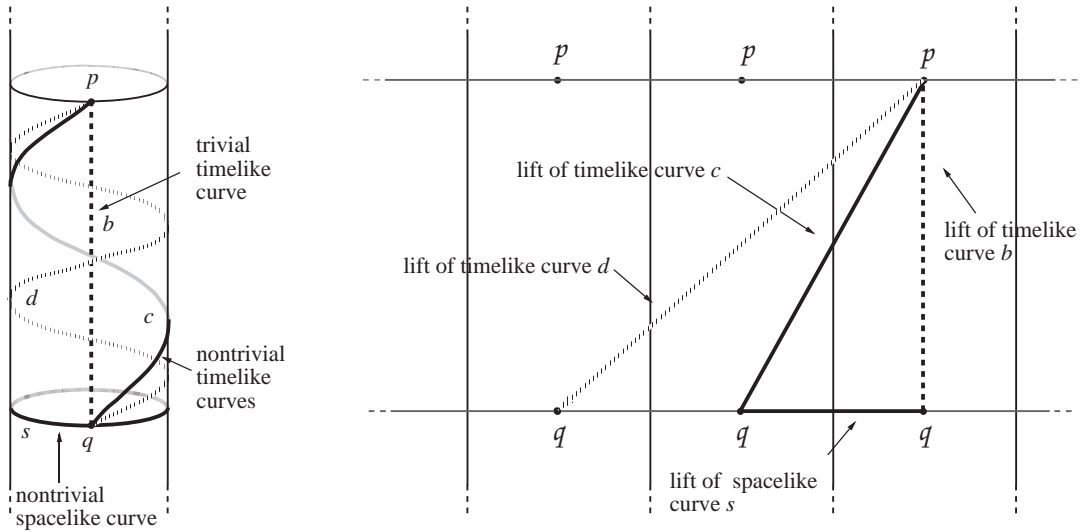
{\baselineskip = 12pt \tenrm \noindent 5. An example of a universal covering
space. On the right is the universal cover of the spacetime illustrated on the
left. All noncontractible curves in the spacetime on the left are unwrapped to contractible
curves in the universal cover.} 
\vskip 0.20in

Although we have illustrated the definition of a universal cover in a particular
example, certain features of this example are generic. Of particular importance is the
fact that noncontractible curves in the original spacetime correspond to
contractible curves connecting copies of the original starting and ending points in the
covering spacetime. This implies that the covering spacetime of an asymptotically
flat spacetime with nontrivial topology always has multiple asymptotically flat
regions even if the original spacetime had only one. Most importantly, any curve
in the original spacetime that traverses the topology necessarily connects two
distinct asymptotic regions in the covering spacetime.

\begin{center}
\vglue .5 cm
\leavevmode
\epsfysize=8cm
\epsfbox{topology2.epsf}
\end{center}
\vskip 0.20in
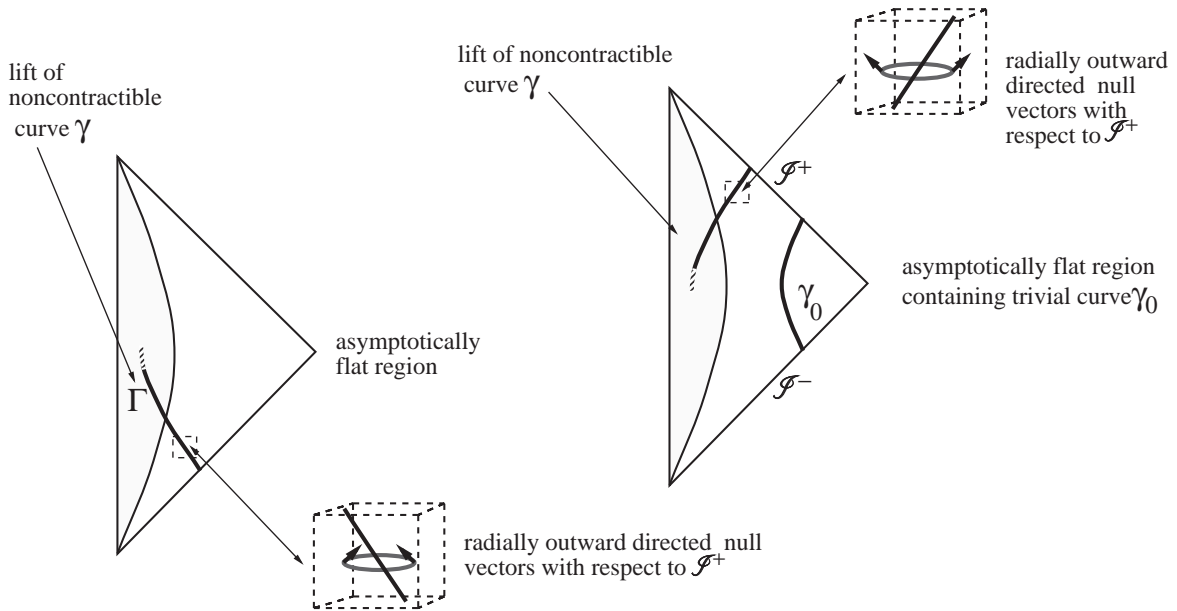
{\baselineskip = 12pt \tenrm 6. An illustration of the proof of the theorem by
contradiction. If the curve $\gamma$ of figure 3 traverses a topological
structure, then it corresponds  to a curve $\Gamma$ in the universal cover
whose beginning is in a different
copy of the asymptotic region than that containing its end and the trivial curve $\gamma_0$.}
\vskip 0.20in

We now have all the tools to discuss the proof of the theorem. The proof is by
contradiction. 
\vskip 14pt

Suppose the theorem is false. Then there is a causal curve
$\gamma$ from $\scri^-$ to $\scri^+$ that passes through the topology and thus is
not deformable to the trivial curve $\gamma_0$. (Recall that figure 3 provides an
 example of such a spacetime.) Now consider the universal
covering space of the spacetime.(A schematic Penrose diagram of the universal covering
space of figure 3 is given in figure 6.) As $\gamma$ passes through the topology,
its corresponding curve $\Gamma$ in the covering space must begin in a different
asymptotic region than that containing $\gamma_0$. In this asymptotic region, the
spacetime is becoming asymptotically flat; therefore, the curve $\Gamma$
intersects arbitrarily large spatial spheres as it approaches the infinite past.
Null curves that can reach $\scri^+$ must be directed inward from these large
spheres as indicated in the inset of figure 6. Therefore photon spheres released
from these surfaces corresponding to these null curves are shrinking in area. Thus
these large spheres near the origin of the curve $\Gamma$ are outer trapped with
respect to an observer at $\scri^+$. However,the assumption  that $\Gamma$
reaches $\scri^+$ means that an observer can see these spheres at $\scri^+$. But this
conclusion contradicts the Lemma! Therefore, our assumption that $\gamma$ is a
causal curve reaching $\scri^+$ must be wrong. Thus there is no
causal curve that passes through the topology and reaches future null
infinity.  As any $\gamma$ not deformable to $\gamma_0$ must correspond to a curve that goes
to another asymptotic region in the universal covering spacetime, we conclude that all curves
reaching future null infinity are deformable to $\gamma_0$ if they are causal. Q.E.D.

\vglue 0.4cm 
\noindent{\elevenbf 4. Discussion} 
\vglue 0.6cm

The consequences of the topological censorship theorem  can be seen by considering a 
physically predictable spacetime with non-Euclidean topology such as a handle 
attached to a plane. Note that this spacetime has
 one asymptotic region. Its universal cover will be a spacetime
with multiple asymptotic regions. Now suppose that an experimenter wishes to
probe the topology of this spacetime and communicate the results of the
measurements to a distant observer near ${\scri}^+$; note that this distant
observer could even be the experimenter herself if she  sends signals or uses 
signals from distant objects  in the spacetime. In order to detect
the handle, the path of some signal must
traverse  the handle and exit to ${\scri}^+$; but this is forbidden by
the theorem.  Only causal paths that do not loop through the handle
can communicate with ${\scri}^+$, and such
causal curves do not detect the existence of non-Euclidean topology.
Thus general relativity prevents one from {\it actively} probing the
topology of spacetime.

So if the topology of the handle cannot be detected, what will the experimenter see?
Note that curvature of spacetime such as that associated with a handle
acts like a mass when viewed from a distant region. Moreover, one cannot
probe the properties of this mass; the topology appears to be behind
a horizon to the experimenter. Thus isolated topological structures 
appear  to be black holes to outside observers, indistinguishable classically
from black holes formed by the collapse of matter. Therefore, if our universe
were full of isolated topological structures, they would appear to us as
black holes.

The theorem was proven for asymptotically flat spacetimes; however,
in cosmology one would like to apply it to spacetimes that do not have the
precise asymptotic behavior of the metric described in the theorem. This is no difficulty
for the case where the scale of the isolated topological structure is small, for
example when it is the size of the solar system (or smaller!) or even of a galactic core.
For these cases our universe is well approximated by an asymptotically flat spacetime.
More generally, note that the key use of the asymptotic behavior of the metric is in
showing  that arbitrarily large spatial spheres are outer trapped in the region from
which $\Gamma$ originates. Intuitively, one expects this behavior to occur in
spacetimes with more general behavior in the asymptotic regions and that the theorem
could be generalized to spacetimes that have metrics that allow arbitrarily large
spatial spheres. Indeed this is the case. Therefore, the topological censorship theorem
applies quite generally to cases of spacetime relevant to cosmology.

Finally, although the sketch of the topological censorship theorem uses the
null energy condition, one can show that it actually can be rigorously proven
for a weaker energy condition, the averaged null energy condition.$^8$ Physically this
energy condition states that the energy \vadjust{\eject} can be negative in small regions so long
as it is positive when averaged over the whole spacetime. This implies that the
topological censorship theorem can be applied not only to spacetimes with classical
matter but may also apply to spacetimes containing certain types of quantum matter.

\vglue 0.6cm 
\noindent{\elevenbf 5. References} 
\vglue 0.4cm

\noindent [1] D.~M.~Witt, {\elevenit Phys. Rev. Lett.} {\elevenbf 57} (1986) 1386  and
to be published.
\vskip 12pt 

\noindent{[2]} J.~Morrow-Jones and D.~M.~Witt, {\elevenit  Phys. Rev. D} {\elevenbf 48 } (1993)  2516.
\vskip 12pt
\noindent{[3]} J.~L.~Friedman, K.~Schleich and D.~M.~Witt, {\elevenit Phys. Rev. Lett.}
{\elevenbf 71} (1993) 1486.
\vskip 12pt
\noindent{[4]} S.~W.~Hawking and G.~F.~R.~Ellis, {\elevenit The Large Scale Structure of
Spacetime}, (Cambridge University Press, Cambridge, 1973) p. 206.
\vskip 12pt
\noindent{[5]} {\elevenit ibid.}, ch. 8.
\vskip 12pt
\noindent{[6]} {\elevenit ibid.}, p. 95.
\vskip 12pt
\noindent{[7]} {\elevenit ibid.}, p. 2.
\vskip 12pt
\noindent{[8]}  A.~Borde, {\elevenit Class. Quantum. Grav.} {\elevenbf 4} (1987) 343.

\end{document}